\documentstyle[12pt,aps,psfig,eqsecnum]{revtex}
\begin{document}

\draft

\title{Hyperscaling in the Broken Symmetry Phase of 
Dyson's Hierarchical Model}

\author{
\vskip50pt 
J. J. Godina\\
Dept. de Fisica, CINVESTAV-IPN , Ap. Post. 14-740, Mexico, D.F. 07000}

\author{Y. Meurice and M. B. Oktay\\ 
{Dept. of Physics and Astr., Univ. of Iowa, Iowa City, Iowa 52242, USA}}


\vskip50pt

\maketitle
\begin{abstract}
\vskip50pt
\hskip140pt
{\bf Abstract}

We use polynomial truncations of the Fourier transform of the local measure
to calculate the connected $q$-point functions of
Dyson's hierarchical model in the broken symmetry
phase. 
We show that accurate values of the connected 1, 2 and 3 point functions
can be obtained at large volume and in a limited range of 
constant external field coupled linearly to the field variable. 
We introduce a new method to obtain the correct infinite volume
and zero external field extrapolations. We extract the leading critical
exponents and show that they obey the scaling and hyperscaling relations
with an accuracy ranging from 
$10^{-5}$ to $5\times 10^{-3}$. We briefly discuss
how to improve the method of calculation.
\end{abstract}
\newpage
\section{Introduction}
\label{sec:intro}
Spontaneous symmetry breaking plays a fundamental role in 
our understanding of the mass 
generation mechanism of the elementary particles. 
One of the simplest field theory model where it is observed is 
scalar theory. Despite its simplicity, there exists no known analytical
method which allows one to elucidate 
quantitatively all the dynamical questions which can be asked 
about scalar field theory in various
dimensions. From a sample of the recent literature on scalar field 
theory, one can see that 
the Monte Carlo method is a popular tool to settle questions 
such as the existence of non-perturbative states \cite{provero}, 
large rescaling of the scalar condensate \cite{consoli} 
or Goldstone mode effects \cite{engels}.

The Monte Carlo method allows us to approach
quantum field theory problems for which 
there are no known reliable series expansions. The main limitations
of the method are the size of the lattice which can be reached
and the fact that the errors usually decrease like $t^{-1/2}$,
where $t$ is the CPU time used for the calculation. 
If, in the next decades, a better knowledge of the fundamental
laws of physics has to rely more and more on precision tests, one 
should complement Monte Carlo methods with new computational 
tools which emphasize numerical accuracy. 

This motivated us to use ``hierarchical approximations''
as a starting point,
since they allow a more easy use of the renormalization group (RG) 
transformation. 
Examples of hierarchical approximations are 
Wilson's approximate recursion formula \cite{wilson} or the 
hierarchical model \cite{dyson}.
In the symmetric phase, we have found \cite{finite} that
polynomial truncations of the 
Fourier transform of the local measure
provide spectacular numerical accuracy, namely, various types of 
errors decrease like
${\rm e}^{-A t^u}$, for some positive constant $A$ of order 1 when $t$
is measured in minutes of CPU time and $0.5\leq u\leq 1$. In particular, 
$t$ only grows as the logarithm of the number of sites 
$L^D$ and the finite-size 
effects decay like $L^{-2}$ when $L$ (the linear size) 
becomes larger than the correlation 
length. This method of polynomial truncations 
was used \cite{gam3} to calculate the critical exponent
$\gamma$ in the symmetric phase for the hierarchical model with 
estimated errors of the order of $10^{-12}$. The result
was confirmed by calculating the largest eigenvalue of the linearized RG
about the accurately determined non-trivial fixed point \cite{wittwer}.

Thanks to the polynomial approximation, 
very accurate information can be encoded in a 
very small set of numbers.
In the symmetric phase, this
approximation is numerically stable when the number of sites becomes 
arbitrarily large and the high-temperature fixed point is reached. 
On the other hand, in the broken symmetry phase, 
numerical instabilities 
appear after a certain number of iterations 
following the bifurcation, and
it is not possible to completely get
rid of the finite size effects with the straightforward procedure
used in the symmetric phase. This issue was briefly discussed in 
section III.E of Ref. \cite{gam3}.

In this paper, we analyze the 
numerical instabilities of the low-temperature phase in a  
quantitative way. We show that in spite of these numerical 
instabilities, it is 
possible to take advantage of the 
iterations for which the low-temperature scaling is observed to 
obtain reliable extrapolations of the magnetization, first to infinite volume 
at non-zero external field and then to zero
external field. We then present a more pratical method of extrapolation 
which 
we apply to calculate
the connected $q$-point functions at zero momentum 
$G_q^c(0)$ for $q=1$, 2 and 3. Finally, we use these calculations to extract 
the leading critical exponents and we check
the hyperscaling relations among these exponents.

The paper is organized as follows. In section II
we show how to construct 
recursively the generating function for the $G_q^c(0)$ when a magnetic
field is introduced.
In section III, we review the scaling 
and hyperscaling relations among the critical exponents and explain
how they should be understood in the case of the hierarchical model.
Hyperscaling\cite{fisher83} usually refers to scaling relations 
involving the dimension explicitly. 
Dyson's hierarchical model has no 
intrinsic dimensionality but rather a continuous free parameter usually 
denoted by $c$ introduced in section II, 
which controls the decay of the 
interactions among blocks of increasing sizes. 
This parameter can be tuned in order 
to insure that a massless field has
scaling properties that can be compared with those of 
nearest neighbor models 
in $D$-dimensions. In the past we have chosen the parametrization 
$c=2^{1-2/D}$, however this is not the only possible one.
In section III. C, we show that a more general parametrization of $c$,
(which includes $\eta$) 
combined with linear arguments yields predictions that are identical 
to the conventional predictions 
obtained from scaling and hyperscaling.
We want to emphasize that the main prediction of the linear theory --
that can be {\it interpreted} as a hyperscaling relation -- 
can be expressed 
in terms of $c$ only and is given in general by Eq. (\ref{eq:gqhm}).
For $c=2^{1/3}$, this general equation together with the accurate 
result of Ref. \cite{gam3} implies
\begin{equation}
\gamma_q=1.29914073\dots \times (5q-6)/4\ ,
\label{eq:test}
\end{equation}
where $\gamma_q$ is the leading exponent corresponding to the 
connected $q$-point function.

We then proceed to verify the predictions of Eq. (\ref{eq:test}) by
doing actual calculations at  various values of 
the inverse temperature $\beta$ near criticality.
This is a rather challenging task because as
one moves away from 
the unstable fixed point, in the low-temperature side, 
rapid oscillations appear in the Fourier transform of the local measure
and the polynomial approximation ultimately breaks down. This
is the cause of the numerical instabilities mentioned above.
As a consequence, a 
relatively small number of iterations can be performed 
with a reasonable accuracy in the low-temperature
phase. This is explained in 
section IV where we also show 
that the number of numerically accurate 
iterations in the low-temperature 
phase scales like the logarithm
of the degree of the polynomial. 
For the calculations discussed later in the paper, we have used 
a polynomial truncation of order 200. With this choice, 
the number of iterations where an 
approximate low-temperature scaling
is observed is slightly larger than 10. Since 
for Dyson's hierarchical model the number of sites is halved after each 
iteration, it means roughly speaking that in correlation length units we
can only reach volumes which are $2^{10}\simeq 10^3$. 
If we use the $D=3$ interpretation of $c=2^{1/3}$, this means 
that the linear size, denoted by $L$, 
which can be reached safely are at most 10 times the
correlation lengths.

Despite this limitation, the magnetization reaches
its infinite volume limit with clearly identifiable  
$L^{-2}$ corrections {\it provided} 
that the external magnetic field is not
too large (otherwise the polynomial approximation breaks down) 
or not too small (otherwise a linear analysis applies and there is no 
spontaneous magnetization). The exact intermediate range of 
the magnetic field 
for which the connected $q$-point functions reach an 
infinite volume limit with the characteristic $L^{-2}$ 
corrections is discussed in section V. In this intermediate range, 
two methods of extrapolation can be used. The first is the standard one 
which consists in 
extrapolating to infinite volume at fixed external field and then
to zero external field. On the other hand, within the intermediate range of 
magnetic field mentioned above,
the magnetization at finite volume
can be fitted very accurately with a straight line which provides an 
extrapolation to zero magnetic field. This extrapolation 
has no physical meaning but it also reaches an infinite volume limit
with $L^{-2}$ corrections when the volume increases. 
This limit coincides with an accuracy of 6 significant figure 
with the limit obtained with the first method; 
in other words within the estimated errors
of the calculation. The second procedure is much more practical because it
does not require any overlap among the acceptable regions of magnetic 
field when the volume increases. The second method will 
be used to calculate the higher point functions.

Proceeding this way, we calculate the connected 
$q$-point functions at zero momentum 
$G_q^c(0)$,
for $q=1$, 2 and 3 and for various values of the inverse temperature $\beta$.
The results are reported in section VI.
The critical exponents are then estimated by using a method discussed in 
Ref. \cite{gam3} where we selected a region of $\beta$ for which the combined 
effects of the errors due to subleading corrections and the numerical 
round-off could be minimized. Using linear fits within this 
limited range of $\beta$, we found 
exponents in agreement with the prediction of hyperscaling given in 
Eq. (\ref{eq:gqhm}) with an accuracy of $ 10^{-5}$ for the 
magnetization, $4\times 10^{-5}$ for the susceptibility and 
$5\times 10^{-3}$ for the 
3-point function. As far as the first two results are concerned, 
the accuracy compares well with the accuracy that can 
usually be reached
with a series analysis or the Monte Carlo method. 
Nevertheless, there is room for  improvement: one should be able
to ``factor out'' the rapid oscillations 
in the Fourier transform of the local measure
and treat them exactly.
This is discussed briefly in the conclusions.

\section{Introduction of a Magnetic Field}
In this section, we describe Dyson's Hierarchical Model
\cite{dyson,sinai} coupled to a constant magnetic field.
All calculations are performed at large but finite volume.
The total number of sites denoted $2^{n_{max}}$.
We label the sites with $n_{max}$
indices $x_{n_{max}}, ... , x_1$, each index being 0 or 1. In order to
visualize this notation, one can divide the $2^{n_{max}}$ sites into
two blocks, each containing $2^{n_{max}-1}$ sites. If $x_{n_{max}}=0$,
the site is in the first box, if $x_{n_{max}} = 1$, the site is in the
second box. Repeating this procedure $n_{max}$ times (for the two boxes,
their respective two sub-boxes , etc.), we obtain an unambiguous
labeling for each of the sites. 

The non-local part of the action (i.e. the ``kinetic term'') 
of Dyson's Hierarchical model reads
\begin{equation}
S_{kin}=
-{\beta\over2}\sum_{n=1}^{n_{max}}({c\over4})^n\sum_{x_{n_{max}},...,x_{n+1}} 
(\sum_{x_n,...,x_1}\phi_{(x_{n_{max}},...,x_1)})^2 \ .
\end{equation}
The index $n$, referred to as the `level of interaction' hereafter,
corresponds to the interaction of the total field in blocks of size $2^n$.
The constant $c$ is a free parameter which describes the way
the non-local interactions decay with the size of the blocks. 
We often  use the parametrization 
\begin{equation}
c=2^{1-2/D}\ ,
\label{eq:convc}
\end{equation}
in order to approximate $D$-dimensional models. 
This question will be discussed later (see Eq. (\ref{eq:cof}) for a 
generalization of Eq. (\ref{eq:convc})).

A constant external source $H$, called ``the magnetic field''
later, is coupled to the total field. This can be represented by 
an additional term in the action 
\begin{equation}
S_H=
-H\sum_{x_{n_{max}},...,x_1}\phi_{(x_{n_{max}},...,x_1)} \ .
\label{eq:magnac}
\end{equation}
However due to the linearity of the coupling, ${\rm e}^{-S_H}$ factorizes into 
local pieces and  this interaction can be absorbed in the local measure.
The field $\phi_{(x_{n_{max}},...,x_1)}$ is integrated over with a local
measure 
\begin{equation}
W_0(\phi,H)\propto W_0(\phi){\rm e}^{H\phi} \ ,
\end{equation}
where $W_0(\phi)$ is the local measure at zero magnetic field.
For simplicity, we use the convention that if the magnetic field
does not appear explicitly in an expression (e. g. , $W_0(\phi)$) 
the quantity 
should be understood at zero magnetic field.
The constant of proportionality refers to the fact that we require 
both $W_0(\phi,H)$ and $W_0(\phi)$ to be normalized as probability 
distributions.
Since we are interested in universal properties, we will use a single 
local measure, namely the Ising measure, $W_0(\phi) = \delta(\phi^2-1)$.
Numerical experiments in Ref. \cite{gam3} show that the universal
properties are very robust under changes in the local measure.

At $H=0$, the recursion relation corresponding to the integration of the 
fields in boxes of size 2, keeping the sum of the 2 fields in each
box constant reads
\begin{equation}
W_{n+1}(\phi) =
{C_{n+1}\over2}e^{{\beta/2}({c/4})^{n+1}\phi^2}\int
d\phi^{'}W_n({(\phi-\phi^{'})\over2})W_n({(\phi+\phi^{'})\over2})\ ,
\label{eq:convol}
\end{equation}
where $C_{n+1}$ is a normalization factor which will be fixed 
in order to obtain a probability distribution.
Introducing the Fourier representation as in Refs.\cite{finite,gam3}
\begin{equation}
W_n(\phi) = \int{dk\over2\pi}e^{ik\phi}\hat{W_n}(k) \ ,
\end{equation}
and a rescaling of the ``source'' $k$ 
by a factor $(c/4)^{1/2}$ at each iteration ,
\begin{equation}
R_n(k) = \hat{W_n}(k(c/4)^{n/2})\ ,
\end{equation}
the recursion relation becomes
\begin{equation}
R_{n+1}(k) = C_{n+1}
exp(-{1\over2}\beta{\partial^2\over\partial
  k^2})(R_n({k(c/4)^{1/2}}))^2.
\label{eq:rec}
\end{equation}
We fix the normalization constant $C_n$ is such way that
$R_n(0)=1$. $R_n(k)$ has then a direct probabilistic interpretation. 
If we call $M_n$ the total field $\sum\phi_x$ inside blocks
of side $2^n$ and $\langle...\rangle_n$ the average calculated without taking into
account the interactions of level strictly larger than $n$, we can
write (at $H=0$)
\begin{equation}
R_n(k) = \sum_{q=0}^{\infty}{(-ik)^{2q}\over(2q)!}{\langle(M_n)^{2q}
\rangle_n(c/4)^{qn}}\ .
\end{equation}

The introduction of the magnetic field is a very simple operation. 
The basic equation reads
\begin{equation}
W_n(\phi,H)\propto W_n(\phi){\rm e}^{H\phi} \ .
\end{equation}
This can be seen in many different ways. One of them is to use Eq. 
(\ref{eq:convol}) and realize that the $\phi'$ drops out of the 
magnetic interactions. Another one consists in realizing that 
due to the linearity, 
one can split Eq. (\ref{eq:magnac}) into sum over boxes of any desired size.
In Fourier transform, this implies that
\begin{equation}
\hat{W}_n(k,H)\propto \hat{W}_n(k+iH)\ .
\end{equation}
The normalization factor is fixed by the condition $\hat{W}_n(0,H)=1$ which 
guarantees that $W_n(\phi,H)$ is a probability distribution and that
$\hat{W}_n(k,H)$ generates the average values of the positive powers
of the total field. More explicitly,
\begin{eqnarray}
\hat{W}_n(k,H)&=&{{ \hat{W}_n(k+iH)}\over{\hat{W}_n(iH)}}\\ \nonumber
& = &\sum_{q=0}^{\infty}{(-ik)^{q}\over q!}{\langle(M_n)^{q}\rangle_{n,H}}\ .
\label{eq:unrescgen}
\end{eqnarray}

From a conceptual point of view, as well as from a practical one,
it is easier to deal with the rescaled quantity $R_n(k)$.  
Near the fixed point of Eq. (\ref{eq:rec}), 
we have the approximate behavior
\begin{equation}
\langle(M_n)^{2q}\rangle_n\propto(4/c)^{qn} \ ,
\label{eq:scaling}
\end{equation}
In terms of the rescaled function, 
we can rewrite Eq. (\ref{eq:unrescgen})
as 
\begin{equation}
{{R_n(k+iH(4/c)^{n/2})}\over{R_n(iH(4/c)^{n/2})}}= 
\sum_{q=0}^{\infty}{(-ik)^{q}\over q!}{\langle(M_n)^{q}
\rangle_{n,H}(c/4)^{qn/2}}\ .
\label{eq:gen}
\end{equation}
The connected Green's functions can be obtained by taking the logarithm
of this generating function. 

\section{About Hyperscaling}
\label{sec:hyp}

\subsection{General Expectations}

The main numerical results obtained in this paper are the calculations of the 
critical exponents corresponding to the singularity of the connected 
$q$-point functions for $q=1,\ 2$ and 3. For definitiness we use the
notation
\begin{equation}
G_q^c(0)\propto(\beta-\beta_c)^{-\gamma_q} \ ,
\end{equation}
for the leading singularities in the low-temperature phase. 
We assume that the reader is familiar with the commonly used 
notations \cite{stanley} for the  critical exponents.
For $q=1$, we have $\gamma_1=-\beta$ which should not be confused
with the inverse temperature. 
After this subsection, we keep using the notation $\beta$ for the 
inverse temperature.
For $q=2$, we have $\gamma_2=\gamma'$.
If one assumes that the scaled magnetization $M/(T-T_c)^\beta$ is
a function of the scaled magnetic field $H/(T-T_c)^\Delta$ only, one obtains
that 
\begin{equation}
\gamma_{q+1}-\gamma_q=\Delta\ ,
\label{eq:gap}
\end{equation}
for any $q$. The exponent $\Delta$ is often called the gap exponent and 
should not be confused with the exponent associated with the subleading 
corrections to the scaling laws.

In general, there exists
7 relations
among the 10 critical exponents $\alpha, \ \alpha '\ , \beta\ ,\gamma\ ,
\gamma '$, $\Delta\ ,\delta\ ,\nu \ ,\nu '\ ,$ and $\eta$,  
in which the dimension of the system does not enter explicitly.
These are the so-called scaling relations \cite{stanley} 
which stimulated the 
development of the RG method. Their explicit form is 
\begin{eqnarray}
\alpha&=&\alpha'\\
\label{eq:gamgam}
\gamma&=&\gamma'\\
\nu&=&\nu'\\
\alpha&+&2\beta+\gamma=2\\
\label{eq:betgam}
\Delta&=&\beta+\gamma\\
\Delta&=&\beta\delta\\
\gamma&=&(2-\eta)\nu \ .
\end{eqnarray}
Eq. (\ref{eq:betgam}) can be seen as 
an obvious version of Eq. (\ref{eq:gap}) for $q=1$, but has also a non-trivial
content summarizing  Eq. (\ref{eq:gap}) for all the higher $q$.

In addition there exists one relation where the dimension enters 
explicitly, for instance: 
\begin{equation}
D\nu = 2-\alpha \ .
\label{eq:nual}
\end{equation}
Other relations may be obtained by combining Eq. (\ref{eq:nual}) with the 
scaling relations. Proceeding this way, we obtain a 
relation of relevance for the rest of the 
discussion, namely
\begin{equation}
\beta = {{(D-2+\eta)}\over{2(2-\eta)}}\gamma \ .
\label{eq:hyper}
\end{equation}
The relations involving the dimension explicitly are usually called
hyperscaling relations \cite{fisher85}. A mechanism leading to 
a possible violation of hyperscaling (dangerous irrelevant variables)
is explained in appendix D of Ref. \cite{fisher83}.
If the 8 relations hold, we are left with only two independents exponents,
for instance $\gamma $ and $\eta$. 

Combining the hyperscaling relation (\ref{eq:hyper}) and the scaling 
relations (\ref{eq:gamgam}) and (\ref{eq:betgam}), we obtain
\begin{eqnarray}
\label{eq:gamq}
\gamma_q&=&\gamma+(q-2)\Delta\\ \nonumber
&=&\gamma\lbrack -2D+q(D+2-\eta)\rbrack /(4-2\eta)
\end{eqnarray}
\subsection{The Hierarchical Model (HT case)}
In the case of the hierarchical model, the exponents $\gamma_q$ 
of the high-temperature (HT) phase (so for $q$ even) can be
estimated by using the linearized RG transformation. 
Since this subsection is the only part of this article where we will 
consider the high-temperature phase, we have not found useful
to introduce special notations for $\gamma_q$ in this phase.
When $\beta_c -\beta$
is small, the linearized RG transformation can be used for approximately
$n^\star$ iterations, with $n^\star $ defined by the relation
\begin{equation}
|\beta-\beta_c|\lambda^{n^\star}=1\ ,
\end{equation}
where $\lambda$ is the largest eigenvalue of the linearized RG transformation.
After the transient behavior has died off and until $n$ reaches the value 
$n^\star$, we are near the fixed point and $R_n(k)$ does not change 
appreciably. Remembering that the field is rescaled by a factor 
$(c/4)^{1\over 2}$ at each iteration (see Eq. (\ref{eq:rec})), 
we obtain the order of
magnitude estimate for $G_q^c(0)$ after $n^\star$ iterations:
\begin{equation}
G_q^c(0)\approx 2^{-n^\star}(4/c)^{qn^\star/2}\ .
\label{eq:bulk}
\end{equation}
For $n$ larger than $n^\star$, the non-linear effects become important.
The actual value of $G_q^c(0)$ may still change by as much as 100 percent,
however the order of magnitude estimate of Eq. (\ref{eq:bulk}) remains valid. 
This transition has been studied in detail in 
Ref.\cite{toy} in a simplified version of the model.
Eliminating $n^\star$ in terms of $\beta_c-\beta$, we obtain
the value of the leading exponents
\begin{equation}
\gamma_q=\gamma \lbrack (q/2){\rm ln} (4/c)-{\rm ln}2\rbrack/{\rm ln}(2/c)\ ,
\label{eq:gqhm}
\end{equation}
with
\begin{equation}
\gamma={\rm ln}(2/c)/{\rm ln}\lambda\ .
\end{equation}
This relationship has been successfully 
tested \cite{finite} in the symmetric phase for
$q=4$ and $4/c=2^{5/3}$. 

\subsection{About Dimensionality}
We will now show that Eq. (\ref{eq:gqhm}) is compatible with the 
general relation of Eq. (\ref{eq:gamq}) provided that
we relate $c$ to a parameter $D$ which can be interpreted as the 
dimension of a nearest neighbor model approximated 
by the hierarchical model. 
We introduce a linear dimension $L$ such that the volume $L^D$ 
is proportional to the total number of sites $2^n$. From
\begin{equation}
L\propto2^{n/D} \ ,
\label{eq:lindim}
\end{equation}
we can in general relate $c$ and $D$ by assuming a scaling of the total field 
\begin{equation}
\langle M_n^{2q}\rangle_n \propto  L^{(D+2-\eta)q}\ ,
\label{eq:dim}
\end{equation}
From comparison with Eq. (\ref{eq:scaling}) this would imply that
\begin{equation}
(4/c)=2^{(D+2-\eta)/D}\ .
\label{eq:cof}
\end{equation}
Substituting in Eq. (\ref{eq:gqhm}), we reobtain the 
general Eq. (\ref{eq:gamq}).

Since in the infinite volume limit, the 
kinetic term is invariant under a RG transformation, 
we have chosen in the past to use Eq. (\ref{eq:cof}) with
$\eta=0$. This is our conventional definition of $c$ given in 
Eq. (\ref{eq:convc}). 
This is the same as saying that when we are near the fixed point,
the total field in a box containing $2^n$ sites, scales with the number of 
sites 
in the same way 
as a massless Gaussian field. This obviously implies that 
in the vicinity of a Gaussian fixed point the total
field scales exactly like a massless Gaussian field in $D$ dimension. 
On the other hand, an interacting massless field will also scale 
like a free one, which is not a bad approximation in $D=3$.
This is an unavoidable feature which will need to be corrected 
when one tries to improve the hierarchical approximation.

We emphasize that this {\it interpretation} has no bearing on the 
validity of the calculations performed. What matters in our calculation 
is the value of $4/c$. In the following, we have used $4/c=2^{5/3}$, which
can be interpreted either 
as $D=3$ and $\eta=0$ or, for instance, as $D=2.97$ and 
$\eta=0.02$. 

\subsection{The Low-Temperature Case}

The extension of the argument for 
odd and even values of $q$ in the broken symmetry phase is somehow 
non-trivial. Since we need to take the infinite volume limit before 
taking the limit of a zero magnetic field, we need 
some understanding of the non-linear behavior. 
Some aspects of the non-linear behavior are discussed in  section 
\ref{sec:extrap}. 
In the following, we will show numerically that Eq. (\ref{eq:gqhm})
holds in good approximation in the broken symmetry phase
for $4/c=2^{5/3}$.
With this choice of $4/c$ and the 
corresponding value of  $\gamma$ calculated in Ref. \cite{gam3}, Eq. 
(\ref{eq:gqhm}) implies Eq. (\ref{eq:test}) given in the introduction.
The verification of this relation for $q$=1, 2 and 3 is the main numerical
result discussed in the following chapters.

\section{Polynomial Truncations}
\label{sec:pol}

In the following we will exclusively consider the case of an Ising 
measure
\begin{equation} 
R_0(k)={\rm cos}(k) \ .
\end{equation}
This restriction is motivated by accurate checks \cite{gam3} 
of universality based on calculations with other measures.
Given that $R_0$ can be expanded into a finite number of eigenfunctions
of ${\rm exp}(-{1\over 2}\beta 
{{\partial ^2} \over 
{\partial k ^2}})$, one can in principle obtain exact expressions for 
the next $R_n(k)$, for instance
\begin{equation} 
R_1(k)={{1+{\rm e}^{\beta c/2}{\rm cos}(k\sqrt{c})}\over{1+{\rm e}^{\beta c/2}}}\ .
\end{equation}
One can in principle repeat this procedure. At each iteration, one obtains 
a superposition of cosines of various frequencies.
For a given numerical value of $c$, $n$ iterations of this exact procedure 
requires to store $2^{n-1}+1$ numerical coefficients. 
The memory size thus scales like $2^n$, while the CPU time
scales like $4^n$.
If $\beta$ differs from $\beta_c$ by $10^{-10}$, one needs at least 80 
iterations in order to eliminate the finite-size effects.
Such a calculation using the exact method 
can be ruled out by practical considerations.

We will thus try to extend
the approximate methods that we have used successfully 
in the symmetric phase \cite{gam3},
where the function $R_n(k)$ was calculated using
finite dimensional approximations \cite{finite} of degree
$l_{max}$:
\begin{equation}
R_n(k) = 1 + a_{n,1}k^2 + a_{n,2}k^4 + ... + a_{n,l_{max}}k^{2l_{max}}\ .
\end{equation}
After each iteration, non-zero coefficients of higher order
($a_{n+1,l_{max+1}}$ etc. ) are obtained, but not taken into account as
a part of the approximation in the next iteration. The recursion 
formula for the $a_{n,m}$ reads \cite{finite} :
\begin{equation}
a_{n+1,m} = {
{\sum_{l=m}^{l_{max}}(\sum_{p+q=l}a_{n,p}a_{n,q}){[(2l)!/(l-m)!(2m)!]}({c/4})^l[-(1/2)\beta]^{l-m}}\over{\sum_{l=0}^{l_{max}}(\sum_{p+q=l}a_{n,p}a_{n,q}){[(2l)!/l!]}{(c/4)^l}[-(1/2)\beta]^l}} \ .
\end{equation}

\begin{figure}
\vskip20pt
\centerline{\psfig{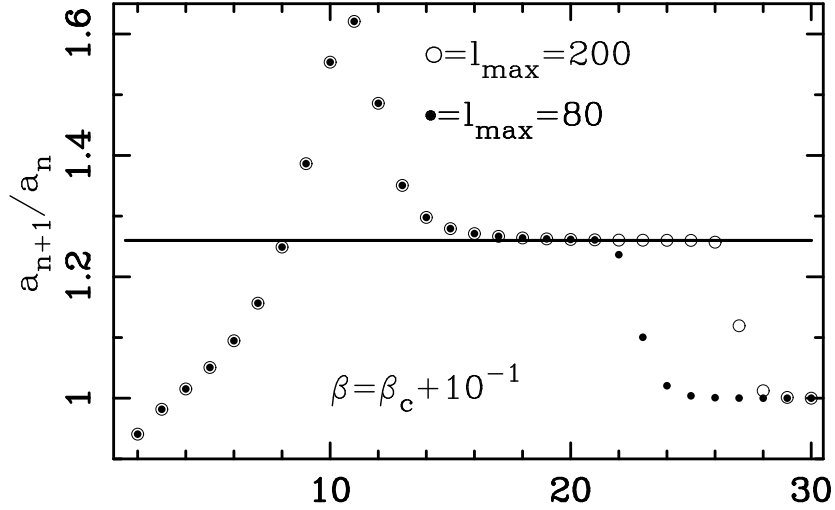}}
\caption{The low-temperature shoulder at $\beta=\beta_c+10^{-1}$
for $\l_{max}=200$ (empty circles)
and $\l_{max}=80$ (filled circles) as a function of $l_{max}$.}
\label{fig:shoulder}
\end{figure}

\begin{figure}
\vskip15pt
\centerline{\psfig{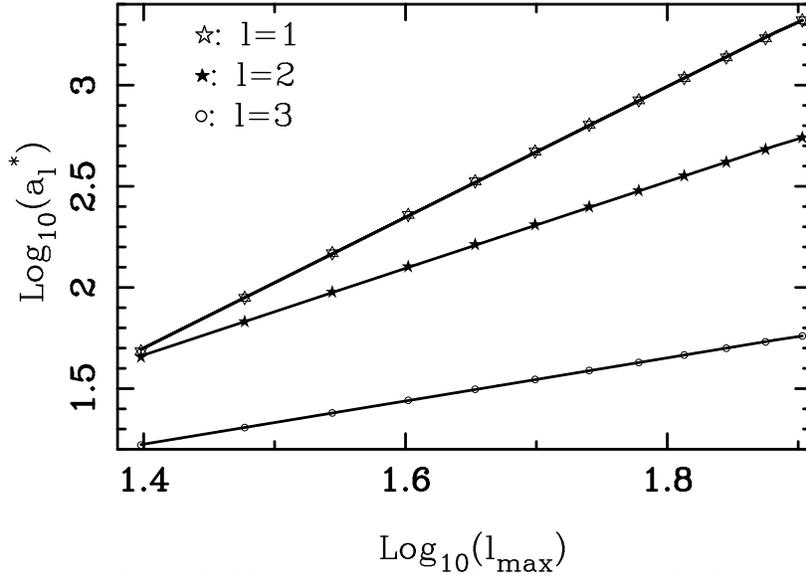}}
\caption{Value of $a_l^\star$ for the ``false'' low-temperature fixed points 
for the Ising case in 3 dimensions for $l=1$ (circles), $l=2$
  (filled stars), $l=3$ (empty stars)}
\label{fig:false}
\end{figure}

The method to identify $\beta_c$ has been discussed in detail in 
Ref.\cite{finite} and consists in finding the bifurcation in the ratio
$a_{n+1,1}/a_{n,1}$. 
In the following, we simply call this quantity ``the ratio''.
If $\beta<\beta_c$, the ratio drops to $c/2$ for $n$ 
large enough. In this case, 
the numerical stability of the infinite volume limit is excellent and allows 
extremely accurate determination of the renormalized quantities.
If $\beta>\beta_c$, the ratio ``jumps'' suddenly a few iterations after
$n^\star$ is reached and stabilizes near the value $c$, corresponding
to the low-temperature scaling. This is seen from 
Eq. (\ref{eq:unrescgen}). Since $\langle M_n^2\rangle_n$ grows like $4^{n}$, as one expects 
in the low-temperature phase, and remembering that there is a rescaling 
of $c/4$ at each iteration, the coefficient of $k^2$ grows like $c^n$.
This implies a ratio equal to $c$.
In our calculation, $c=1.25992\dots$. Unfortunately, the number of iterations
where the low-temperature scaling is observed is rather small. Subsequently,
the ratio drops back to 1. As we shall explain at length, this is an
effect of the polynomial truncation. The length of the ``shoulder'' were the 
low-temperature scaling is observed increases if we increase $l_{max}$.
This situation is illustrated in Figure \ref{fig:shoulder}.
No matter how large $l_{max}$
is, for $n$ large enough, the ratio eventually drops back to 1. 
This reflects the existence of a {\it stable} fixed point
for the {\it truncated} recursion formula.
The values $a_l^\star$ of $a_l$ at this fixed point for various $l_{max}$
are  shown in Figure \ref{fig:false}.  
We see a clear evidence for a dependence of the form
\begin{equation}
a_l^\star\propto(l_{max})^l\ .
\end{equation}
This means that the stable fixed point is an effect of the 
polynomial truncations and has no counterpart in the original model.

It is possible to evaluate the value of $n$ for which the low-temperature
shoulder ends. A detailed study shows that for $n$ large enough, we 
have in good approximation
\begin{equation}
R_n(k)\simeq {\rm cos}({\mathcal{M}}c^{n/2}k)\ ,
\label{eq:apprn}
\end{equation}
where $\mathcal{M}$ is the magnetization density in the infinite
volume limit. 
If we assume that $R_n(k)$ is exactly as in 
Eq. (\ref{eq:apprn}), then we can use 
the basic recursion formula (\ref{eq:rec}) in  order to
obtain the corresponding $R_{n+1}(k)$. Using 
$2\times {\rm cos}^2(x)=1+{\rm cos}(2x)$, we can 
reexpress  $(R_n({k(c/4)^{1/2}}))^2$ as a superposition
of eigenfunctions
of the one-dimensional Laplacian. 
When the exponential of the Laplacian in Eq. (\ref{eq:rec})
acts on the non-constant modes it becomes exp($\beta{\mathcal{M}}^2
c^{n+1}/2$). In the polynomial truncation of the recursion relation,
this exponential is replaced by $l_{max}$ terms of its Taylor 
expansion. This approximation is valid if the argument of the 
exponential is much smaller than $l_{max}$.
Consequently, we obtain that the polynomial truncation certainly
breaks down if $n$ is larger than $n_b$ such that
\begin{equation}
n_b+1\simeq \lbrack{\rm ln}(2/\beta)-{\rm ln}(
{\mathcal{M}}^2)+{\rm ln}(l_{max})\rbrack/{\rm ln}c \ .
\label{eq:nb}
\end{equation}
If the estimate of Eq. (\ref{eq:bulk}) extends to the low-temperature 
phase, one realizes that the second term of (\ref{eq:nb}) 
is roughly $n^\star$ while the third term stands for the length of the 
peak and the shoulder. Plugging the approximate 
values 1.1 for $\beta$ and 0.7 for
$\mathcal{M}$ (see section \ref{sec:extrap}), 
we obtain $n_b=23$ for $l_{max}=80$ and  
$n_b=27$ for $l_{max}=200$. A quick glance at Figure \ref{fig:shoulder},
shows that these estimates coincide with the first drastic drops of the 
low-temperature shoulder.

One can in principle extend indefinitely the low-temperature shoulder
by increasing $l_{max}$. However, the CPU time $t$ necessary for $n$ 
iterations 
of a quadratic map in dimension $l_{max}$ grows like
\begin{equation}
t\propto n(l_{max})^2 \ .
\label{eq:cpu}
\end{equation}
As we will show in section \ref{sec:extrap}, 
the finite-size effects on $G_q^c(0)$ are of
the order $(c/2)^{n_s}$ where $n_s$ is the number of points on the shoulder.
This behavior has been 
demonstrated \cite{finite}
in the high-temperature phase and we will see later that
it also applies in the low-temperature phase. From the previous discussion 
$n_s\approx {\rm ln}l_{max}/{\rm ln}c$. This implies that the finite-size
effects $\mathcal{E}$ are of the order
\begin{equation}
{\mathcal{E}}\propto (l_{max})^{\ln (c/2)/\ln c}\ .
\end{equation}
Using Eq. (\ref{eq:cpu}) and the value of $c$ expressed in terms of $D$
according to Eq. (\ref{eq:cof}) with $\eta=0$, we obtain
\begin{equation}
{\mathcal{E}}\propto t^{-1/(D-2)}\ .
\end{equation}
In particular, for the value $4/c=2^{5/3}$ used hereafter, 
the errors decrease like $t^{-1}$.
Consequently, we should try to modify the method in such a way that
the rapidly oscillating part of $R_n(k)$ is treated without polynomial
approximations. This possibility is presently under investigation. 
One can nevertheless obtain results with an accuracy competing with 
existing methods by using the finite data on the short shoulder in 
order to extrapolate to the infinite volume limit result.
This procedure is made possible by the rather regular way the 
renormalized quantities approach this limit.

\section{The Extrapolation to Infinite Volume}
\label{sec:extrap}
\subsection{Preliminary Remarks}

There is no spontaneous magnetization at finite 
volume. This well-known statement can be understood directly from
Eq. (\ref{eq:gen}). As explained at the beginning of section 
\ref{sec:pol}, at finite $n$, $R_n(k)$ is simply a superposition
of cosines with finite positive coefficients provided that $\beta$ is real.
However if $\beta$ is complex,
these coefficients have singularities. This comes from 
the normalization factor, needed when we impose the condition $R_n(0)=1$, 
which has zeroes in the complex plane.
The behavior of these zeroes has been studied in Ref. \cite{osc} for $n$
between 6 and 12. As the volume increases, these zeroes ``pinch'' the 
critical point. However at finite $n$, there are no zeroes on the real
axis. In conclusion, at real $\beta$ and finite $n$,  $R_n(k)$
is an analytical function of $k$. For any given $n$, we can always take 
the magnetic field $H$ small enough in order to have 
\begin{equation}
|H(4/c)^{n/2}|\ll 1 \ .
\end{equation}
If we express $c$ in terms of the linear dimension using Eqs. (\ref{eq:convc})
and (\ref{eq:lindim}) this translates into
\begin{equation}
|H|\ll L^{-(D+2)/2} \ .
\end{equation}

Given the analyticity of $R_n(k)$, one can then use Eq. (\ref{eq:gen}) in the 
linear approximation. In this limit,
\begin{equation}
\langle M_n\rangle _n\simeq -2a_{n,1}H(4/c)^n \ ,
\label{eq:lin}
\end{equation}
and the magnetization vanishes linearly with the magnetic field.

On the other hand, for any non-zero $H$, no matter how small its
absolute value is, one can always find a $n$ large enough to have 
$|H(4/c)^{n/2}|\gg 1$. The non-linear effects are then important and 
Eq. (\ref{eq:lin}) does not apply. In addition it is assumed (and will 
be verified explicitly later) that when such a $n$ are reached, the value
of the $G_q^c(0)$ stabilizes at an exponential rate. One can then, {\it first}
extrapolate at infinite volume for a given magnetic field, and {\it then}
reduce the magnetic field in order to extrapolate a sequence of 
infinite volume
limits with decreasing magnetic field, 
towards zero magnetic field. Again, this procedure requires some 
knowledge about the way the second limit in reached. In the case considered
here (one scalar component), the limit is reached by linear extrapolation.
In systems with more components, the Nambu-Goldstone modes create a square 
root behavior \cite{wilczek} 
\begin{equation}
M(T<T_c,H>0)=M(T,0^+)+CH^{1/2} \ .
\end{equation}
which has been observed for $O(4)$ models  
using Monte Carlo simulations \cite{engels}. We now discuss the application
of the procedure outlined above in the simplest case.

\subsection{Calculation of the magnetization}

\begin{figure}
\vskip20pt
\centerline{\psfig{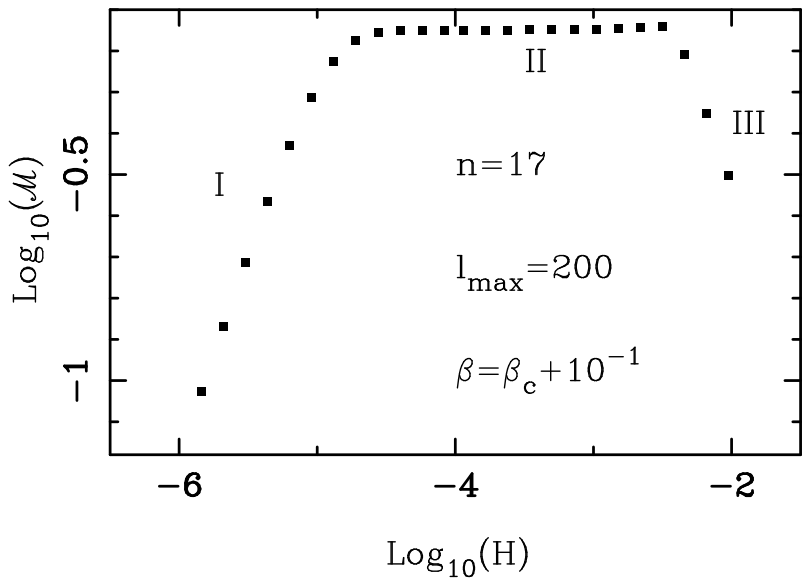}}
\caption{$log_{10}({\mathcal{M}})$ versus $\log_{10}(H)$ at $n=17$ 
for $\beta=\beta_c+10^{-1}$.}
\label{fig:magnetization}
\end{figure}

In this subsection we discuss the calculation of the 
infinite volume limit of the magnetization. 
The magnetization density at finite volume 
is defined as
\begin{eqnarray}
{\mathcal{M}}_n(H) &=& {\langle M_n\rangle _{n,H}\over 2^n} \ .
\end{eqnarray}
We call it ``the magnetization'' when no 
confusion is possible.
For definiteness, we have chosen a special value $\beta=\beta_c+10^{-1}$
and calculated the magnetization by plugging numerical values of $H$
in Eq. (\ref{eq:gen}) and expanding to first order in $k$. The results are
shown in Fig. \ref{fig:magnetization} for $n=17$ and $l_{max}=200$.
As one can see, we have three different regions. The first one (I)
is the region where the linear approximation 
described above applies. For the 
example considered here, the linearization condition 
$|H(4/c)^{n/2}|\ll 1$ translates into ${\rm log}_{10}(H)\ll -4.3$. This is 
consistent with the fact that the linear behavior is observed below -5.
The third part (III), is the region where the polynomial approximation 
breaks down. Given the approximate form given in Eq. (\ref{eq:apprn}),
this should certainly happen when $|H(4/c)^{n/2}|\approx l_{max}$.
This means  ${\rm log}_{10}(H)\approx -2.0$ in our example. On the figure,
one sees that for ${\rm log}_{10}(H)\approx -2.4$, the magnetization drops
suddenly instead of reaching its asymptotic value at large $H$, namely 
${\mathcal{M}}=1$. Finally, the intermediate region (II) is the one which
contains the information we are interested in. 

\begin{figure}
\vskip20pt
\centerline{\psfig{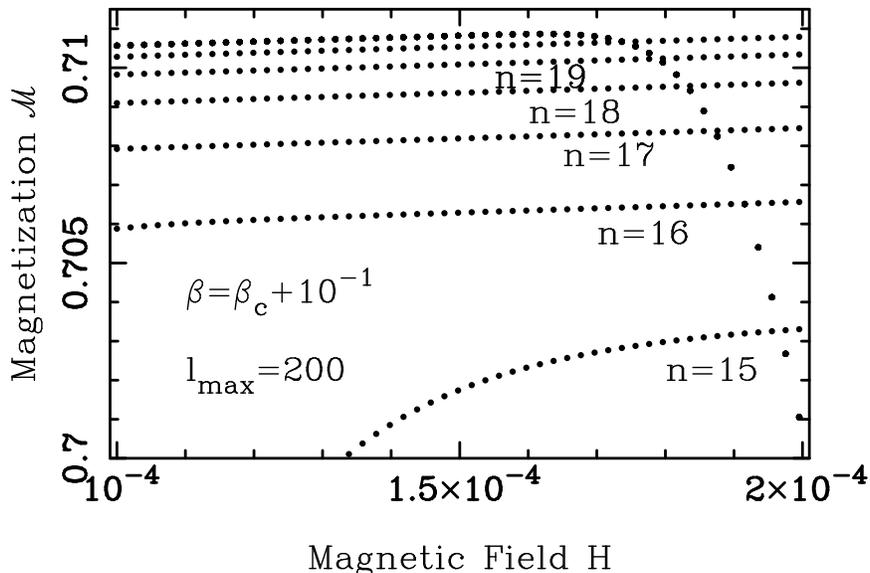}}
\caption{The magnetization versus the magnetic field for $n=15$ 
(lower set of point) to 21 (upper set of points on the left of the 
figure) for
$l_{max}=200$ and $\beta=\beta_c+10^{-1}$. } 
\label{fig:linear200}
\end{figure}

As advertized, we will first take the infinite volume limit of the
magnetization at non-zero magnetic field 
and then extrapolate to zero magnetic field. 
We need to understand how the second region shown
in Fig. \ref{fig:magnetization} changes with $n$. From the above discussion,
region II is roughly given by the range of magnetic field
\begin{equation}
-(n/2){\rm log}_{10} (4/c)<{\rm log}_{10} (H)<{\rm log}_{10}( l_{max})
- (n/2){\rm log}_{10}(4/c) \ .
\end{equation}
In the log scale of Fig. \ref{fig:magnetization}, the width of region II
is at most ${\rm log}_{10}( l_{max})$ which is approximately 2.3 in our sample 
calculation. 
Region II shifts by 
$-(1/2){\rm log}_{10} (4/c)$, approximately 0.25 in our sample 
calculation, at each iteration. In addition, the whole graph moves slightly
up at each iteration in a way which is better seen using a linear scale
as in Fig. \ref{fig:linear200}.
As one can see, the regions II of seven successive iterations do not  
overlap. Consequently 1.5 is a more realistic estimate than the previously 
quoted bound 2.3 for the 
average width of region II. 

\begin{figure}
\vskip20pt
\centerline{\psfig{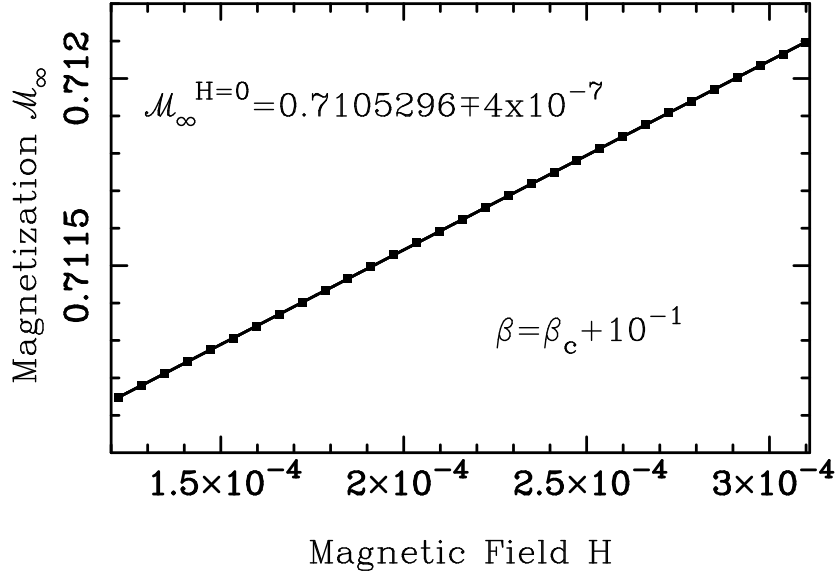}}
\vskip20pt
\caption{${\mathcal{M}}_\infty$ extrapolated  
from  the
data for  $n$= 17, 18, 19 
and 20, versus the magnetic field for
$l_{max}=200$ and $\beta=\beta_c+10^{-1}$.}
\label{fig:infinitefit}
\end{figure}

The fewer iterations we use to extrapolate to infinite volume,
the broader the range of the magnetic field can be.
We have compared 5 sets of 4 iterations well on the low-temperature shoulder
starting from the set (17, 18, 19, 20) up to the set (21, 22, 23, 24).
From our experience in the symmetric phase \cite{finite} we have 
assumed that the finite-size effects could be parametrized as 
\begin{eqnarray}
\label{eq:infvol}
{\mathcal{M}}_n &=& {\mathcal{M}}_\infty - A\times B^n \ . 
\end{eqnarray}
This relation implies that
\begin{eqnarray}
\log_{10}({\mathcal{M}}_{n+1}-{\mathcal{M}}_{n}) & =& \tilde{A} 
+ n\times\log_{10}(B)\ ,
\end{eqnarray}
where $ \tilde{A} =  \log_{10}(A) + \log_{10}(1-B)$. 
The value $ \tilde{A}$ and $\log_{10}(B)$ can be obtained from linear fits.
For four successive iterations, this  will give us 
3-point fits for the infinite volume limit 
at fixed value of $H\neq0$. In all fits performed, we found $B\simeq 0.63$,
which is compatible with the $(c/2)^n$ decay of the finite size effects found 
in the symmetric phase \cite{finite}. In terms of the linear dimension $L$ 
introduced in Eq. (\ref{eq:lindim}), this corresponds to finite-size effects 
decaying like $L^{-2}$.
If the parametrization of Eq. (\ref{eq:infvol}) was exact, the value of 
$ {\mathcal{M}}_n +A\times B^n$ would be independent of $n$ and equal
to ${\mathcal{M}}_\infty$. In practice, variations 
slightly smaller than $10^{-6}$
are observed. We have thus taken an average over these values in order to 
estimate ${\mathcal{M}}_\infty$ at fixed $H$. The results for the 
first set are shown in Fig. \ref{fig:infinitefit}
for various values of $H$. The linear behavior 
allows an easy extrapolation to $H=0$.

We have repeated this procedure for the four other set of successive
values defined previously and obtained the $H=0$ extrapolations:
\vspace{10mm}
\begin{quasitable}
\begin{tabular}{cccc}
Set & ${\mathcal{M}}_{\infty}^{H=0}$ \\
\tableline
1 & 0.7105296 \\ 
2 & 0.7105349 \\
3 & 0.7105376\\
4 & 0.7105380 \\
5 & 0.7105382  
\end{tabular}
\end{quasitable}
\vspace{10mm}
Averaging over these five values, we obtain 
\begin{eqnarray}
\label{eq:estim}
{\mathcal{M}}_{\infty}^{H=0}&=&0.710536\mp3 \times10^{-6} \ .
\end{eqnarray}
It may be argued that the values coming from sets involving larger 
values of $n$ are better estimates because the finite size effects 
are smaller for those sets.  

We have repeated this type of calculation with sets of 5
successive iterations and a correspondingly narrower range of 
magnetic field and found results compatible with the estimate 
given by Eq. (\ref{eq:estim}).

\subsection{The Susceptibility}

We now consider the calculation of the connected susceptibility
(two-point function). By using the previous notation, we can express it as
\begin{eqnarray}
\chi_n(H) &=& {\langle M_n \rangle_{n,H}^{2} - \langle M_{n}^{2} \rangle_{n,H} 
\over 2^n}\ 
\nonumber \\
&=& {(b_{1}^{2}-2\times b_{2}) \over 2^n}\ ,
\end{eqnarray}
where 
\begin{equation}
{{R_n(k+iH(4/c)^{n/2})}\over{R_n(iH(4/c)^{n/2})}}= 
\sum_{q=0}^{\infty}b_q{k^{q}}\ .
\end{equation}
The dependence on $H$ of the $b_n$ is implicit.

\begin{figure}
\vskip20pt
\centerline{\psfig{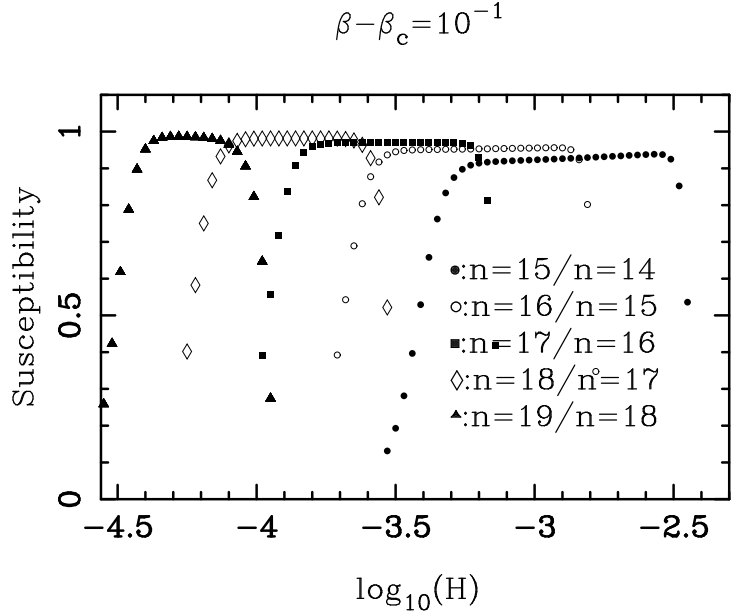}}
\vskip20pt
\caption{The successive ratios of the susceptibility for 
$\beta=\beta_c+10^{-1}$.}
\label{susmag}
\end{figure}

In order to extrapolate the susceptibility to infinite $n$,  
one has to determine 
the range of the magnetic field for which  
the scaling $\langle M_n \rangle_{n,H}^{2} - \langle M_{n,H}^{2} 
\rangle_{n}\propto2^n$ holds. 
When this is the case, the ratio $\chi_{n+1}/\chi_n \simeq 1$.
The range of values of the magnetic field for which this scaling is observed
is analogous to ``region II''  introduced in the previous subsection, and
we will use the same terminology here.    
The ratios of the susceptibility at successive $n$
are shown for various values of $H$ in Fig. \ref{susmag}.

\begin{figure}
\vskip20pt
\centerline{\psfig{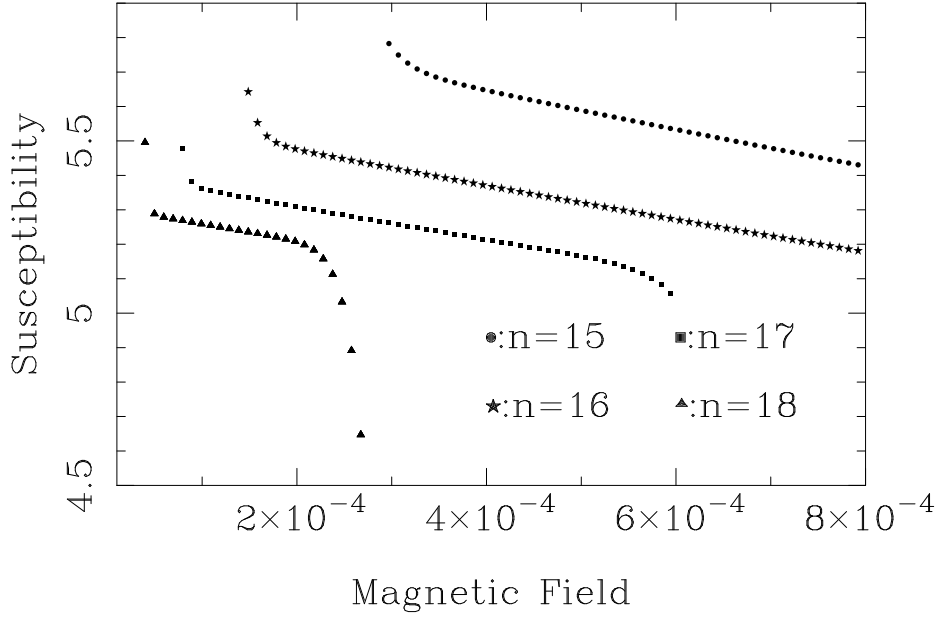}}
\vskip20pt
\caption{The susceptibility $\chi_n$ versus the magnetic field $H$ at different
$n$ for $\beta=\beta_c+10^{-1}$.}
\label{susep}
\end{figure}

One observes that the range where the desired scaling is observed 
shrinks when $n$ increases. For each successive iteration, the ratio of 
susceptibility has a ``upside-down U''-shape. The values of $H$ for which
the ratio starts dropping on the left are equally spaced and can be 
determined by linearization as before. On the other side of the 
upside-down U, dropping values 
of the ratio signal the breakdown of the polynomial truncation. 
This occurs at smaller values of $H$ than for the magnetization, making
the region II smaller.
A theoretical estimate of the lower value of $H$ for which this happens
requires a more refined parametrization than the one given 
in Eq. (\ref{eq:apprn}). 
In order to get a controllable extrapolation, we need at
least 4 successive values of $\chi_n$ (to get at least 3-point fits
for the logarithm of the differences). This is unfortunately impossible: the 
region II of three successive upside-down U have no overlap as one can 
see from Fig. \ref{susmag}. Similar results are obtained by plotting 
the susceptibility versus the magnetic field as shown in Fig. \ref{susep}.
One sees that the regions II (where the susceptibility can approximately
be fitted by a line which extrapolates to a non-zero value at zero $H$) do not 
overlap for 4 consecutive iterations. 

\subsection{An Alternate Method}

\begin{figure} 
\vskip20pt
\centerline{\psfig{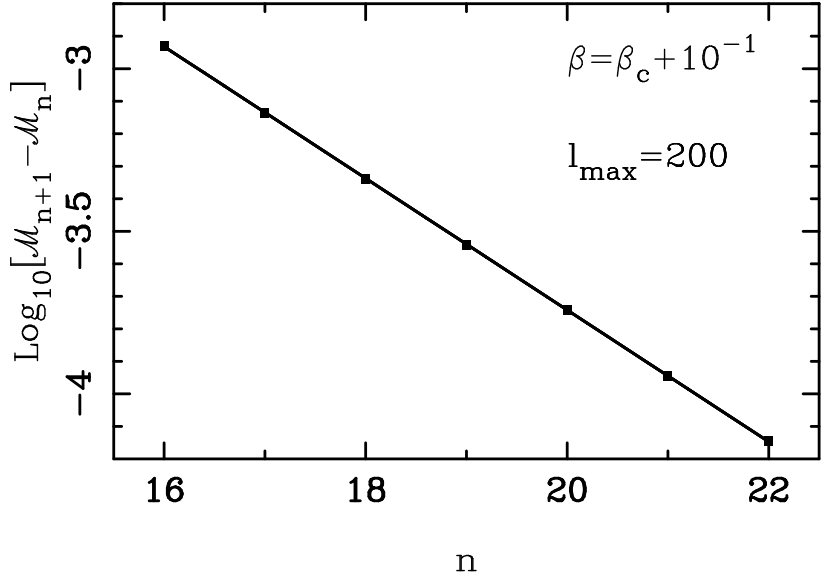}}
\vskip20pt
\caption{$\log_{10}({\mathcal{M}}_{n+1}^{H\rightarrow 0 }-
{\mathcal{M}}_{n}^{H\rightarrow 0})$ 
versus $n$ for
$l_{max}=200$ and $\beta=\beta_c+10^{-1}$ .}
\label{fig:7point}
\end{figure}

In the previous discussion, we have observed a linear behavior for the 
region II of the magnetization and the susceptibility. This linear behavior
can be used to obtain extrapolations to non-zero values of these quantities
at zero magnetic field. These values have no physical interpretation. 
We denote them by ${\mathcal{M}}_{n}^{ H\rightarrow 0}$, the arrow 
indicating that the quantity is a mathematical extrapolation and {\it not}
``the spontaneous magnetization at finite volume''. 
They reach an asymptotic value at an exponential suppressed rate 
when $n$ increases, just as in 
Eq. (\ref{eq:infvol}). This is illustrated in Fig. \ref{fig:7point}. 
Using a linear fit to fix the unknown parameters $A$ and $B$ 
in Eq. \ref{eq:infvol}, and 
averaging the ${\mathcal{M}}_{n}^{H\rightarrow 0}+AB^n$ over $n$, we obtain 
\begin{eqnarray}
{\mathcal{M}}_{\infty}^{H=0} &=& 0.710537\mp 2\times10^{-6}\ ,
\end{eqnarray}
which is consistent with the result obtained with the standard 
method.

Roughly speaking, the lines of region II move 
parallel to each other when $n$ increases and it is 
approximately equivalent to first extrapolate to zero $H$, using
the linear behavior in  region II, and then to 
infinite $n$ rather than the contrary. 
If the two limits coincide, 
the second method has a definite practical advantage:
all we need is a small part of region II for each $n$, no matter if it 
overlaps or not with the region II for other $n$. So in general, it allows
to use more iterations to get better quality extrapolations.
The fact that the values of the magnetization obtained with 
the two methods coincide with 5 significant digits is a strong 
indication that the two procedure are equivalent. 
For the susceptibility and higher point 
functions, we do not have an independent check, since the alternate 
method is the only one available. However, we were able to make 
consistency checks such as the fact that the slope of the straight line
used for the zero magnetic field extrapolation of the $q-$point function
coincides with the $q+1$-point function.

We can repeat the same steps for the 3-point function. The 3-point
function is given by 
\begin{eqnarray}
G_3^c &=& {M_3-3M_1M_2+2M_{1}^{3}\over 2^n} \nonumber \\
&=& {6b_3-6b_1b_2-2b_{1}^{3}\over 2^n} \ ,
\end{eqnarray}
where the dependence on $H$ is implicit.
As shown in Ref. 
\cite{glimm}, $G_3^c<0 $ for $H\leq0$. 
Due to the additional subtraction, the range where the proper low-temperature 
scaling is observed is smaller than for the suceptibility.
It is not possible to repeat the same steps for the
4-point function which is given by,
\begin{eqnarray}
G_4^c &=& M_4-3M_{2}^{2}-4M_1M_3+12M_{1}^{2}-6M_{1}^{4} \ ,
\end{eqnarray}
and involves one more subtraction.

\section{Estimation of the Exponents}

We have used the method described in the previous section to calculate the 
value of the connected $q$-point functions at value of $\beta$ approaching
$\beta_c$ from above with equal spacings on a logarithmic scale. 
For reference, the numerical values are given in the table below. 
\begin{quasitable}
\begin{tabular}{cccc} \\ 
\tableline
$-\log_{10}(\beta-\beta_c)$ & $G_{1}^{c}(0)$ & $G_{2}^{c}(0)$ & 
$G_{3}^{c}$(0) \\
\tableline
1 & $0.710537$ & $5.1449$ & $-452$ \\ 
2 & $0.372929$ & $147.75$ & $-4.83\times 10^{6}$ \\
3 & $0.181173$ & $3270.0$ & $-4.42\times10^{8} $ \\
4 & $8.64639\times10^{-2}$ & $67534$ & $-3.82\times10^{11}$ \\
5 & $4.10479\times10^{-2}$ & $1.3628\times10^6$ & $-3.23\times10^{14}$ \\
6 & $1.94518\times10^{-2}$ & $2.7276\times10^7$ & $-2.72\times10^{17}$ \\
7 & $9.21183\times10^{-3}$ & $5.4411\times10^8$ & $-2.28\times10^{20}$ \\
8 & $4.36138\times10^{-3}$ & $1.0842\times10^{10}$ & $-1.91\times10^{23}$ \\
9 & $2.06473\times10^{-3}$ & $2.1596\times10^{11}$ & $-1.60\times10^{26}$ \\
10 & $9.77434\times10^{-4}$ & $4.3010\times10^{12}$ & $-1.34\times10^{29}$ \\
11 & $4.62716\times10^{-4}$ & $8.5641\times10^{13}$ & $-1.11\times10^{32}$ \\
12 & $2.19084\times10^{-4}$ & $1.7042\times10^{15}$ & $-9.40\times10^{34}$ \\
\tableline
\end{tabular}
\end{quasitable}

The estimated errors on the values quoted above are of order 1 in the 
last digit for the first lines of the table and slowly increase when
one moves down the table. For the last lines, the effects of the round-off 
errors become sizable. Otherwise, the errors are mainly due to the 
extrapolation procedure. We have checked that the numerical values of
the quantities at finite $H$ had reaches their asymptotic values 
(well within the accuracy of the final result) as 
a function of $l_{max}$. 

\begin{figure}
\vskip15pt
\centerline{\psfig{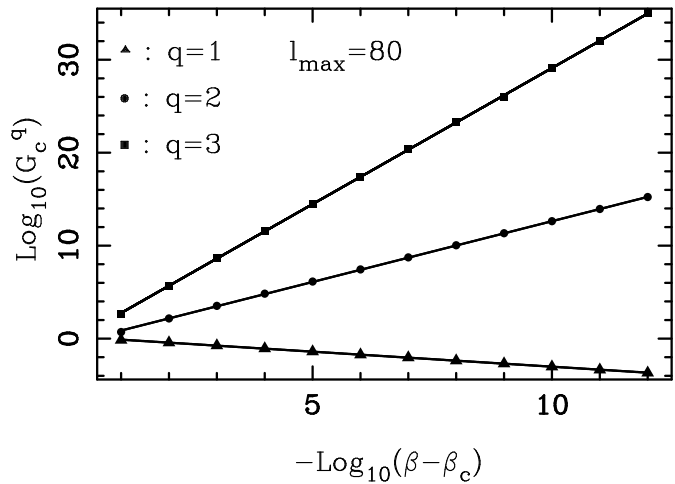}}
\vskip15pt
\caption{${\rm log}_{10}(G_c^q(0))$ versus ${\rm log}_{10}(\beta-\beta_c)$ for 
$q$=1, 2 and 3.}
\label{fig:123ex}
\end{figure}

The results are displayed in Fig. \ref{fig:123ex} in a log-log plot.
The departure from the linear behavior is not visible on this figure.
In the symmetric phase, we know \cite{gam3} 
that the relative strength of the subleading 
corrections is approximately -0.57$(\beta_c-\beta)^{0.43}$. It is likely 
that a similar behavior should be present in the low-temperature phase. 
Consequently, taking into account the data on the left part of 
Fig. \ref{fig:123ex}
will distort the value of the exponents. 
On the other hand getting too close to criticality generates large 
numerical errors. 
Using a linear fit of the data 
starting with the 5-th point and ending with 
the 10-th point, we obtain the value of the exponents 
\begin{eqnarray}
\gamma_1&=&-0.3247 \nonumber \\
\gamma_2&=&1.2997\\
\gamma_3&=&2.9237 \nonumber \ .
\end{eqnarray}
This can be compared with the predictions from scaling and  hyperscaling
given by Eq. (\ref{eq:test}) and amount numerically to:
\begin{eqnarray}
\label{eq:pred}
\gamma_1&=&-0.324785 \nonumber \\
\gamma_2&=&1.2991407\\
\gamma_3&=&2.923066 \nonumber .
\end{eqnarray}
Better estimates can be obtained by using 
the method developed in Ref. \cite{gam3}, where it was found that 
the combined effects of the two types of errors are minimized 
for $10^{-10}<|\beta_c-\beta|<10^{-9}$. 
This allowed estimates of $\gamma$ (in the symmetric phase) 
with errors of order $3\times10^{-5}$ (compared to more accurate estimates).
Using 10 values between 
$10^{-9}$ and $10^{-10}$ with equal spacing on a logarithmic scale,
we obtain here
\begin{eqnarray}
\label{eq:best}
\gamma_1 &=& -0.324775\pm 2\times 10^{-5}\nonumber \\
\gamma_2 &=& 1.29918\pm 10^{-4}\\
\gamma_3 &=& 2.928\pm 10^{-2} \nonumber .
\end{eqnarray}
The errors due to the subleading corrections and the round-off errors
are approximately of the same order  in this region of temperature \cite{gam3}.
The errors due to the subleading corrections are larger for larger values
of $|\beta_c-\beta|$ while the numerical errors are larger for
smaller values
of $|\beta_c-\beta|$ . 
We have estimated the errors due to the subleading corrections 
by performing the same calculation between 
$10^{-8}$ and $10^{-9}$. The errors bars quoted above reflect the differences
with the exponents obtained in this second region.
\section{Conclusions}
One sees clearly that our best estimates of the critical exponents 
(Eq. (\ref{eq:best}))
are fully 
compatible with the predictions of hyperscaling 
(Eq. (\ref{eq:pred})). The differences between the predicted and 
calculated values are $10^{-5}$ for $\gamma_1$, 
$4\times 10^{-5}$ for $\gamma_2$ and
$5\times 10^{-3}$ for $\gamma_3$. They fall well within the estimated errors. 
Since hyperscaling is a reasonable expectation, this also 
shows that the non-standard 
extrapolation method that we have used is reliable. As far as $\gamma_1$ and 
$\gamma_2$ are concerned, the errors bars are smaller than what can usually
be reached using a series analysis or Monte Carlo simulation. Our result
for $\gamma_1$ is also compatible with the result 0.325 obtained in Ref. 
\cite{baker} for the hierarchical model (for $\sigma/d=2/3$ with their 
notations) using the integral formula.

One could in principle improve the accuracy of these calculations by increasing
the size of the polynomial truncation. However, the efficiency of this 
procedure (errors decreasing like the inverse of the CPU time) is not 
compatible with our long term objectives (errors decreasing exponentially).
The main obstruction to keep using the polynomial truncation is that the 
generating function $R_n(k)$ starts oscillating rapidly in the low-temperature
phase making the approximation of the exponential of the Laplacian by
a sum
inaccurate. It is thus important to obtain an approximate parametrization 
of  $R_n(k)$ in terms of eigenfunctions of the Laplacian. A step in this 
direction is made by the parametrization of Eq. (\ref{eq:apprn}). This 
approximate analytical form needs to be improved in order to include the 
connected 2-point and higher point functions in terms of an exponential function. 
This possibility is presently under investigation.
 


This research was supported in part by the Department of Energy
under Contract No. FG02-91ER40664.

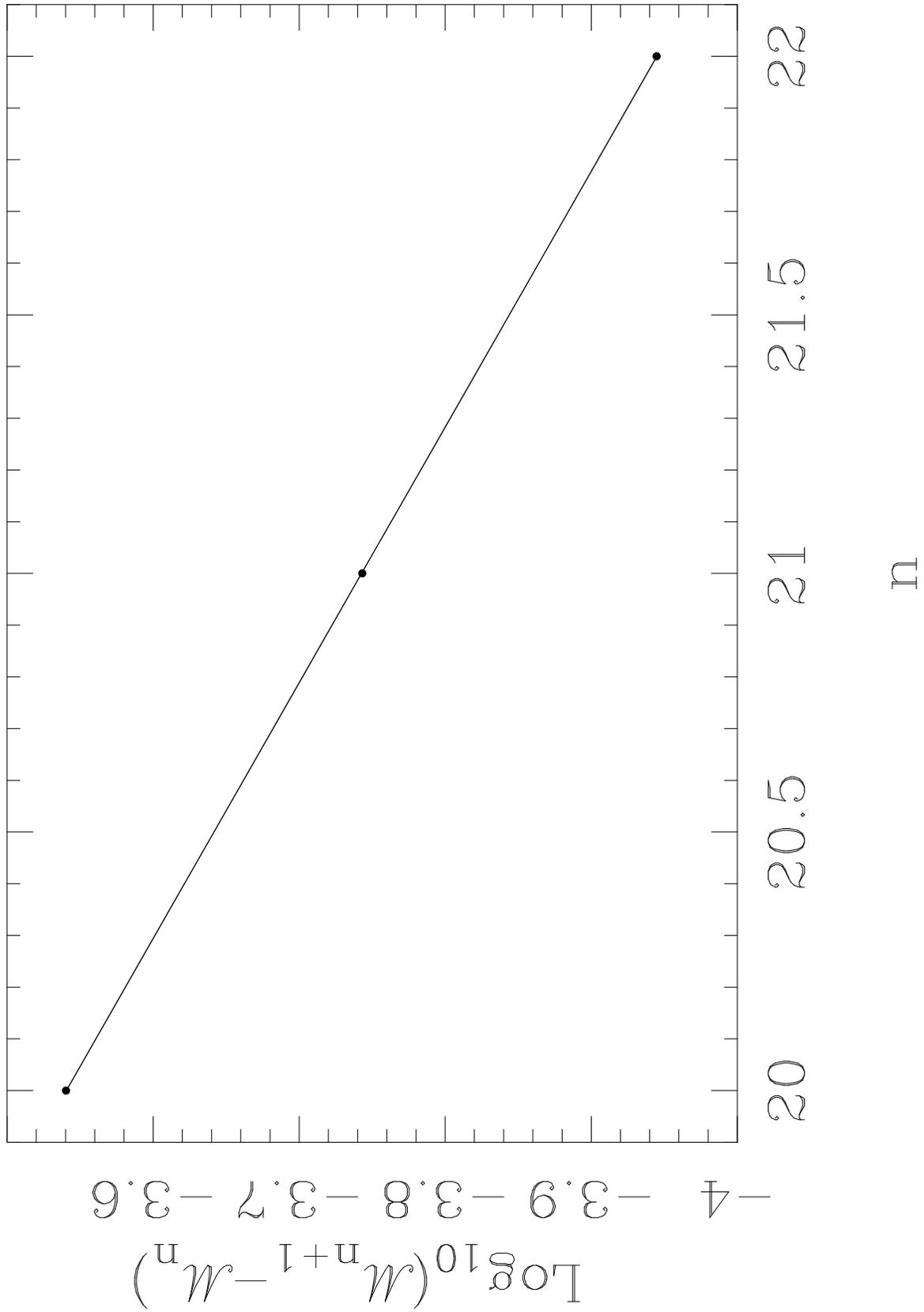
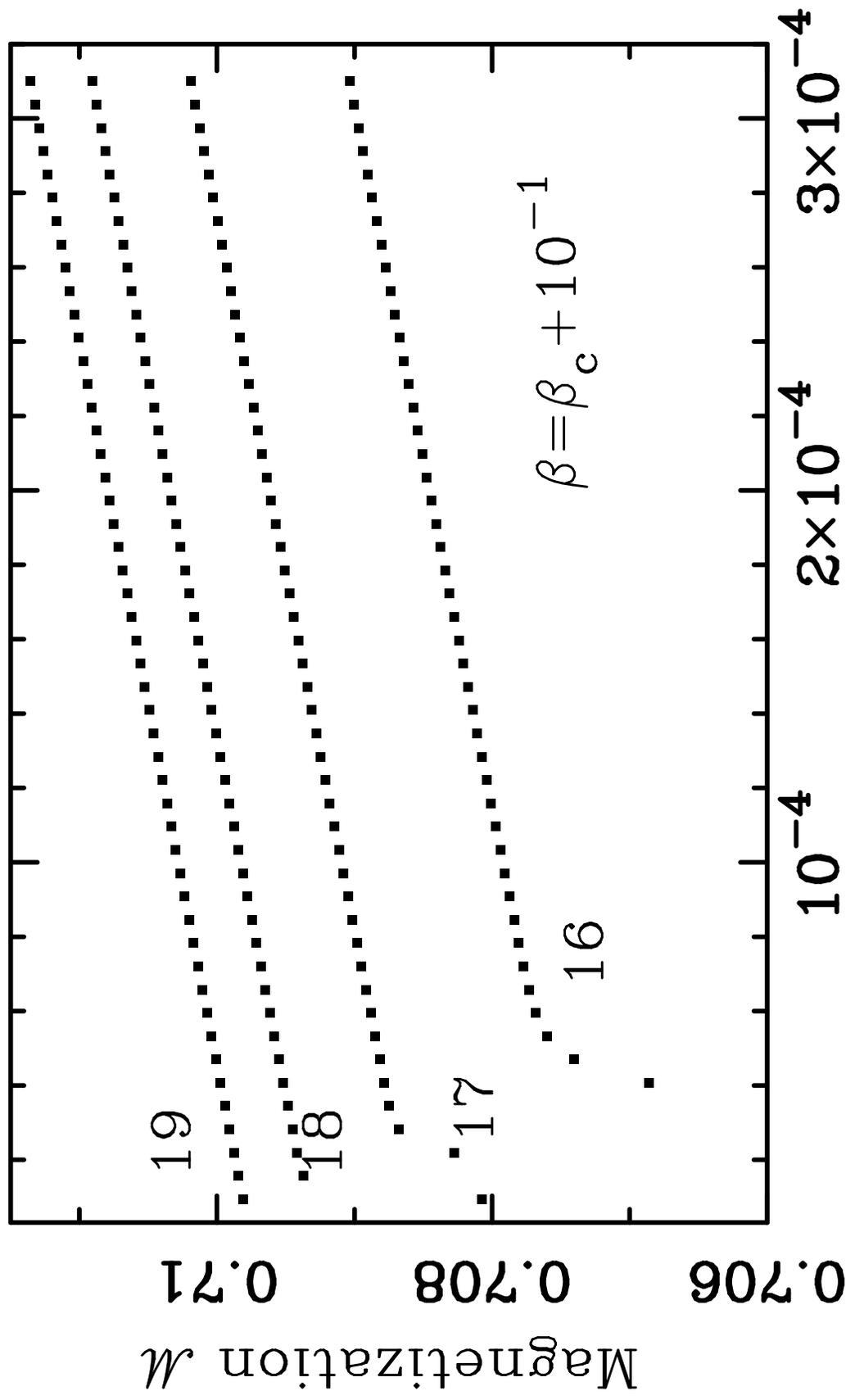
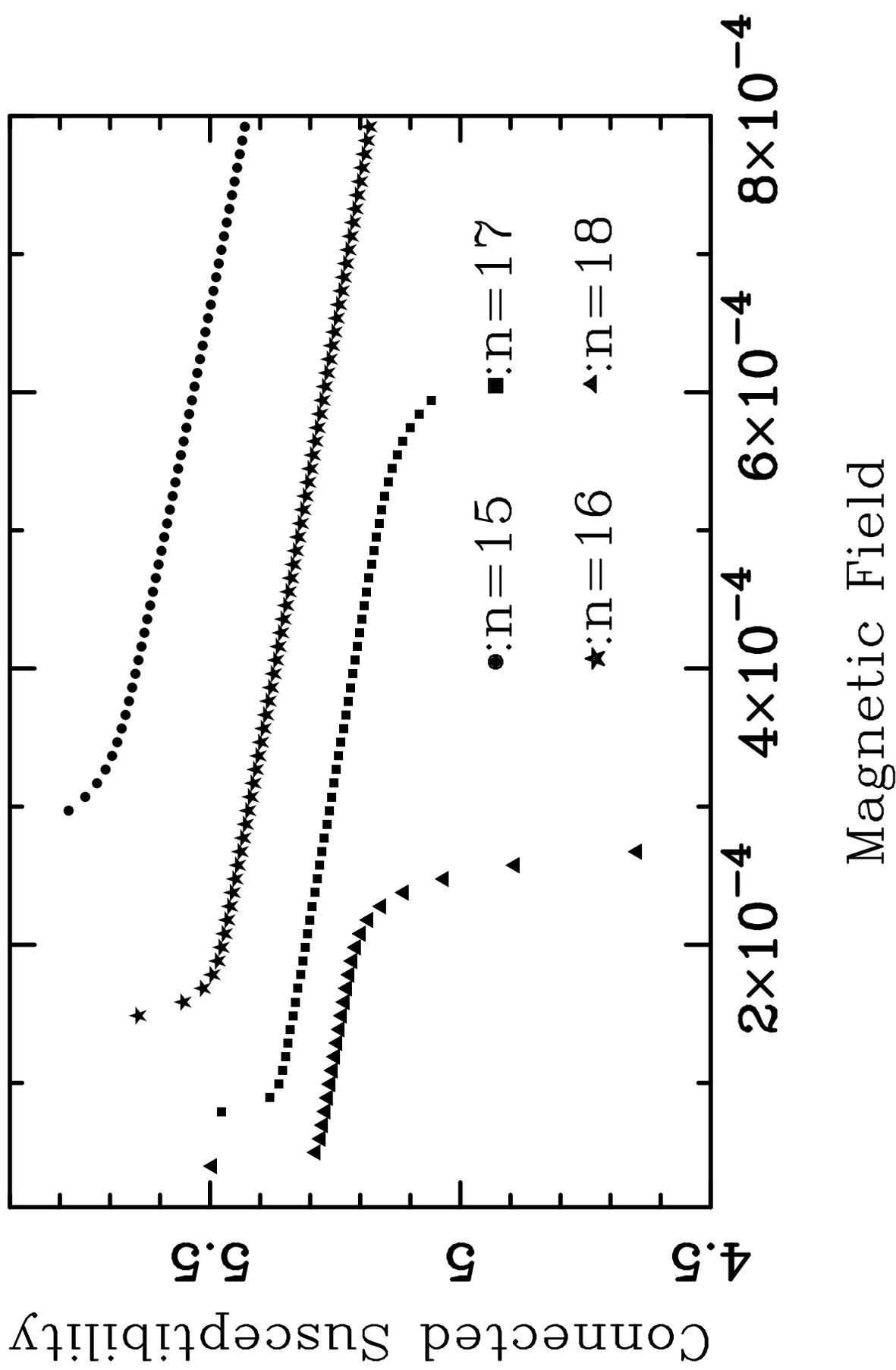
\end{document}